\documentstyle[aps,epsf,preprint]{revtex}

\begin{document}

\title{Nonlinear dynamics in superlattices driven by high
frequency ac-fields}

\author{J.\ Devis and E.\ Diez}

\address{GISC, Departamento de Matem\'aticas, Universidad Carlos
III, E-28911 Legan\'{e}s, Madrid, Spain}

\author{V.\ Bellani}

\address{INFM-Dipartimento di Fisica ``A.\ Volta", Universit\'{a}
di Pavia,  I-27100 Pavia, Italy}

\date{\today}

\maketitle

\begin{abstract}

We investigate the dynamical processes taking place in
nanodevices driven by high-frequency electromagnetic fields.
We want to elucidate the role of different mechanisms
that could lead to loss of quantum coherence. Our results show
how the dephasing effects of disorder that destroy  after some periods
coherent oscillations, such as Rabi oscillations, can be overestimated  
if we do not consider the electron-electron interactions
that can reduce dramatically the decoherence effects
of the structural imperfections. Experimental conditions for the
 observation of the predicted effects are discussed.  
\end{abstract}


\newpage

Recent advances in laser technology make possible to drive semiconductor
nanostructures with intense coherent ac-dc fields.  This opens new
research fields in time-dependent transport in mesoscopic
systems~\cite{Cundiff} and puts forward the basis for a new
generation of ultra-high speed devices. Following these results
several works have been devoted to the analysis of the effects of 
time-dependent fields on the transport properties of resonant 
heterostructures~\cite{Diez1,Diez2,Diez3},  and to exploit the
application in ultrafast optical technology: high-speed optical switches,
coherent control of excitons, etc.~\cite{Citrin1,Citrin2}.

Within this context, we are interested in the decoherence
processes producing the observed fast dephasing of  coherence
phenomena in semiconductor superlattices (SL's)
and more specifically in 
the interplay between the growth imperfections
(disorder) and many-body effects as the  electron-electron (e-e)
interaction.
The interplay between the effects of disorder and many-body effects
on electronic properties is a long-standing problem in solid-state physics.
Probably one of the most promising way to gain insight into this intricate
problem is to combine the actual state-of-the-art of the Molecular Beam Epitaxy (MBE), which allow us to grew samples with monolayer perfection
and consequently with well-characterized disorder, with coherent
oscillations that are extremely sensitive to imperfections and nonlinear effects.

The oscillations of a two level system between the ground and excited
states in the presence of a strong resonant driving field, often called
transient nutation or Rabi oscillation (RO), are  discussed in
textbooks as a topic of time-dependent perturbation theory.
Consider a two state system with ground state energy $E_0$ and excited
state $E_1$ in the presence of a harmonic perturbation.  If the
frequency of the perturbation matches roughly the spacing between the
two levels, the system undergoes oscillations with a frequency $\Omega_R$
which is much smaller than the excitation frequency
$\omega_{\mathrm{ac}}$.  This Rabi frequency depends on the mismatch
$\delta \omega \equiv (E_1 - E_0)/\hbar - \omega_{\mathrm{ac}}$ between
the level spacing and the excitation frequency, and on the
matrix element $F_{10}$ of the perturbation $\Omega_R = \left( \delta
\omega^{2} + |F_{10}|^2/\hbar^{2} \right)^{1/2}$.  If we start with the
system initially in the ground state, transitions between the ground and
the excited state will occur with a period $T_R = 2 \pi/\Omega_R$ as
time evolves.

Semiconductor SL's present Bloch minibands with
several states each one; thus it is not clear whether they can be
correctly described as a a pure two state system.  We should also take
into account the presence of imperfections introduced
during growth processes and  scattering mechanism as e-e interactions.
Interface roughness appearing during growth in {\em actual} SL's depends
critically on the growth conditions.  For instance,
protrusions of one semiconductor into the other cause in-plane disorder
and break translational invariance parallel to the layers.  To describe
local excess or defect of monolayers, we allow the quantum well widths
to fluctuate uniformly around the nominal values; this can be seen as
substituting the nominal sharp width by an {\em average} along the
parallel plane of the interface imperfections.  Our approximation is
valid whenever the mean-free-path of electrons is much smaller than the
in-plane average size of protrusions as electrons only {\em see} {\em
micro\/}-quantum-wells with small area and uniform thickness.  Therefore, in the following we will take the width of the
$n$th quantum well to be $a(1+W\epsilon_n)$, and the width of the $n$th
barrier as $b(1+W\epsilon_n)$ where $W$ is a positive parameter
measuring the maximum fluctuation, $\epsilon_n$'s are distributed
according to a uniform probability distribution, ${\cal{P}}(\epsilon_n)=1$ if
$|\epsilon_n|<1/2$ and zero otherwise, $a$ is the nominal quantum well
width and $b$ is the nominal quantum barrier width.
Even with its rather satisfactory degree of success, many-body
calculations have difficulties that, in some cases, may complicate the
interpretation of the underlying physical processes.
Presilla et {\em al}.~\cite{Presilla} sugges\-ted a new treatment of
these interactions that, loosely speaking, could be regarded as
similar to Hartree-Fock and other self-consistent
techniques, which substitute many-body interactions by a nonlinear
effective potential. More recently~\cite{Diez4} we have proposed a new model
where the nonlinear interaction is driven by a local field instead of
the mean-field approach used by Presilla and co-workers.\cite{Presilla}.

Here we present a new model that solves self-consistently the Poisson and
the time-dependent
Sch\"odinger  equations, and we compare the results with a 
local treatment based in the non-linear Schr\"odinger equation. 


The band structure and the potential at flat band is computed by 
using a finite-element
method.  The eigenstate $j$ of the band $i$ with
eigen\-ener\-gy $E_{i}^{(j)}$ is denoted as $\psi_{i}^{(j)}(x)$.  A good
choice for the initial wave packet is provided by using a linear
combination of the eigenstates belonging to the first miniband.  For the
sake of clarity we have selected as the initial wave packet $\Psi(x,0) =
\psi_{i}^{(j)}(x)$, although we have checked that this assumption can be
dropped without changing our conclusions.  The subsequent time evolution
of the wave packet $\Psi(x,t)$ is calculated numerically by means of an
implicit integration schema designed for consider time-dependent 
fields~\cite{Diez5}. 

The envelope-functions for the electron wavepacket satisfies the
following quantum evolution equation
\begin{equation}
\label{e1}
i\hbar\frac{\partial \Psi(x,t)}{\partial t} = \left[-\frac{\hbar^{2}}{2m^{*}}
\frac{d^{2}}{dx^{2}} + V_{NL}(x,t)\right]\Psi(x,t),
\end{equation}
where $x$ is the coordinate in the growth direction of the SL. We consider
 two approaches to the nonlinear potential $V_{NL}(x,t)$ in Eq.(~\ref{e1}).
 On the one hand we take into account the model described, 
 in other context, in~\cite{Diez4} for our problem, where $V_{NL}(x,t)$ is
 \begin{equation}
 \label{e2}
 V_{NL}(x,t) = V(x) - eF_{AC}x \sin(\omega_{AC}t) +
 \alpha_{loc}|\Psi(x,t)|^{2},
 \end{equation}
and $V(x)$ is the potential at flat-band, $F_{AC}$ and $\omega_{AC}$ are
the strength and the frequency of the ac field respectively, and all
the nonlinear physics is contained in the coefficient $\alpha_{loc}$, 
which we discuss below.
There are several factors that configure the medium nonlinear response
to the tunneling electron. We want to consider only the repulsive 
electron-electron Coulomb interactions, which should enter the effective
potential with a positive nonlinearity, i.e., the energy is increased by
local charge accumulations, leading to a positive sign for $\alpha_{loc}$.

On the other hand, we have considered a different approach by solving
self-consistently the Schr\"odinger and Poisson equations obtaining
a Hartree-like potential. In this context, the non-linear potential is,
\begin{equation}
\vspace*{-0.25cm}
\label{e3}
V_{NL}(x,t) = V(x) - eF_{AC}x \sin(\omega_{AC}t) +
\alpha_{self}V_{H}(x,t),
\end{equation}

where now $V_{H}$ 
it is obtained by solving the Poisson equation for the density of charge
$|\Psi(x,t)|^{2}$, and $\alpha_{self}$ is the
coupling parameter.


We present here results for a SL with $10$ periods of $100\,$\AA\ GaAs
and $50\,$\AA\ Ga$_{0.7}$Al$_{0.3}$As with conduction-band offset  $300\,$meV and
$m^{*}=0.067m$, $m$ being the free electron mass. 
To illustrate the effects of the nonlinear coupling we show
in Fig.~\ref{1} 
the conduction-band profile  for a perfect SL ($W=0$) 
at $t=0.4$ (lower) and $1.2\,$ps (upper)
when the ac field is tuned to the resonant frequency 
$\omega_{\mathrm{ac}} = \omega_{\mathrm{res}} \sim 24\,$THz,
for (a) the linear case and modeling  the e-e interaction with (b)  
the self-consistent method ($\alpha_{self} = 10^{-3}$) and  with (c)
the local model ($\alpha_{loc} = 10$). 
To show the effects of the interface
roughness we plot in  Fig.~\ref{2} 
the probability of finding an electron,
initially si\-tua\-ted in
$\psi_{0}^{(5)}(x)$, in the state $\psi_{1}^{(5)}(x)$ as a function of time 
when the ac field is tuned to the resonant frequency
$\omega_{\mathrm{ac}} = \omega_{\mathrm{res}} \sim 24\,$THz with (a) $W=0$
(perfect SL) and (b) $W=0.03$ (imperfections around one monolayer).
These results suggest the existence of a
characteristic
scattering time $\tau_{\mathrm{dis}}$ related to the amount of disorder
in the sample, after which RO's are destroyed by disorder.
In Fig.~\ref{3} we plot the probability of finding an electron,
initially situated in
$\psi_{0}^{(5)}(x)$, in the state $\psi_{1}^{(5)}(x)$ as a function of time
when the ac field is tuned to the resonant frequency, for
different values of the nonlinearity coupling (a) $\alpha_{self} = 5\, \times 10^{-5}$, 
(b) $10^{-4}$, (c) $5\, \times 10^{-4}$ and (d) $10^{-3}$. The results for the
local model are very similar. When we compare this picture with Fig.~\ref{2}
we see the process of vanishing of the RO's are completely different.
In the second case, the effects are the same for any time, then we could not
speak about a dephasing time, apparently we only modified the electronic
structure and then we are decreasing the resonant coupling between the
external ac-field and the Bloch bands.
Nevertheless the main goal of this work is to show how the nonlinearity
effects can reduce the dephasing effects introduced by the 
growth imperfections. In Fig.~\ref{4} we plot  the
occupation probability of the state $\psi_{1}^{(5)}(x)$ as a function of time
considering imperfections about one monolayer ($W=0.03$) for (a) the linear
case and considering together with the imperfections the e-e interaction
with the self-consistent model (b) $\alpha_{self} = 10^{-4}$ and with the local one (c) $\alpha_{loc} = 5$.
We can see clearly how nonlinearity prevents the dephasing effects introduced
by the imperfections allowing  the observation of Rabi oscillations
during larger coherence times. These theoretical results are completely 
consistent
with  recent experiments in transport properties of intentional disordered
superlattices with doped and undoped superlattices\cite{Koch,Bellani}, 
where is showed than the  Coulomb
interactions could be  the responsible of the suppression of disorder effects
leading to quasimetallic behavior at low temperatures when 
the doping of the samples is increased.


In summary, we have shown how the dephasing effects of disorder
are dramatically reduced when we consider the e-e interaction.
We have studied two different models to introduce the non-linear
interaction and the results are very similar.
Our results shows that it is possible to enlarge
the dephasing times and,  consequently 
the number of periods of coherence oscillations of electrons in SL's.
In semiconductor  heterostructures this can be done by 
increasing the doping or with very intense laser excitation fields.
It goes without saying that to develop new devices for THz science it is
crucial to understand
how to control and enlarged the coherence times. We think that the 
nonlinear effects could be the key to solve this problem.
Further work along these lines is currently in progress.

The authors would like to thank Francisco Dom\'{\i}nguez-Adame for
helpful discussions and critical reading of the manuscript. Also
we thank to
I.\ Bossi, Rafael G\'omez-Alcal\'a, Claudio Andreani and Gennady Berman, 
for very valuable conversations. J.D. and E.D. thanks the
Dipartimento di Fisica ``A.\ Volta'' of the Universit\'a di Pavia
for hospitality during a stay when part of this work was done.
Work in Madrid was supported by CAM under 
Project No.~07N/0034/1998, and in Pavia by the
INFM Network ``Fisica e Tecnologia dei Semiconduttori III-V''.
J.D. and E.D. also acknowledges partial support from
Fundaci\'on Universidad Carlos III (Spain) and INFM (Italy).

\begin{figure}
\caption{Conduction-band profile.}
\label{1}
\end{figure}

\begin{figure}
\caption{The probability of finding the electron
in the state $\psi_{1}^{(5)}(x)$
as a function of time for different amounts of disorder.}
\label{2}
\end{figure}

\begin{figure}
\caption{The probability of finding the electron in the state
$\psi_{1}^{(5)}(x)$
as a function of time for different values of the nonlinear coupling.}
\label{3}
\end{figure}

\begin{figure}
\caption{The probability of finding the electron in the state
$\psi_{1}^{(5)}(x)$
as a function of time 
considering the interplay between imperfections and nonlinearities.}
\label{4}
\end{figure}


\begin{thebibliography}{99}
\bibitem{Diez5} Details about the numerical procedure and algorithms
will be given elsewhere.
\end{thebibliography}
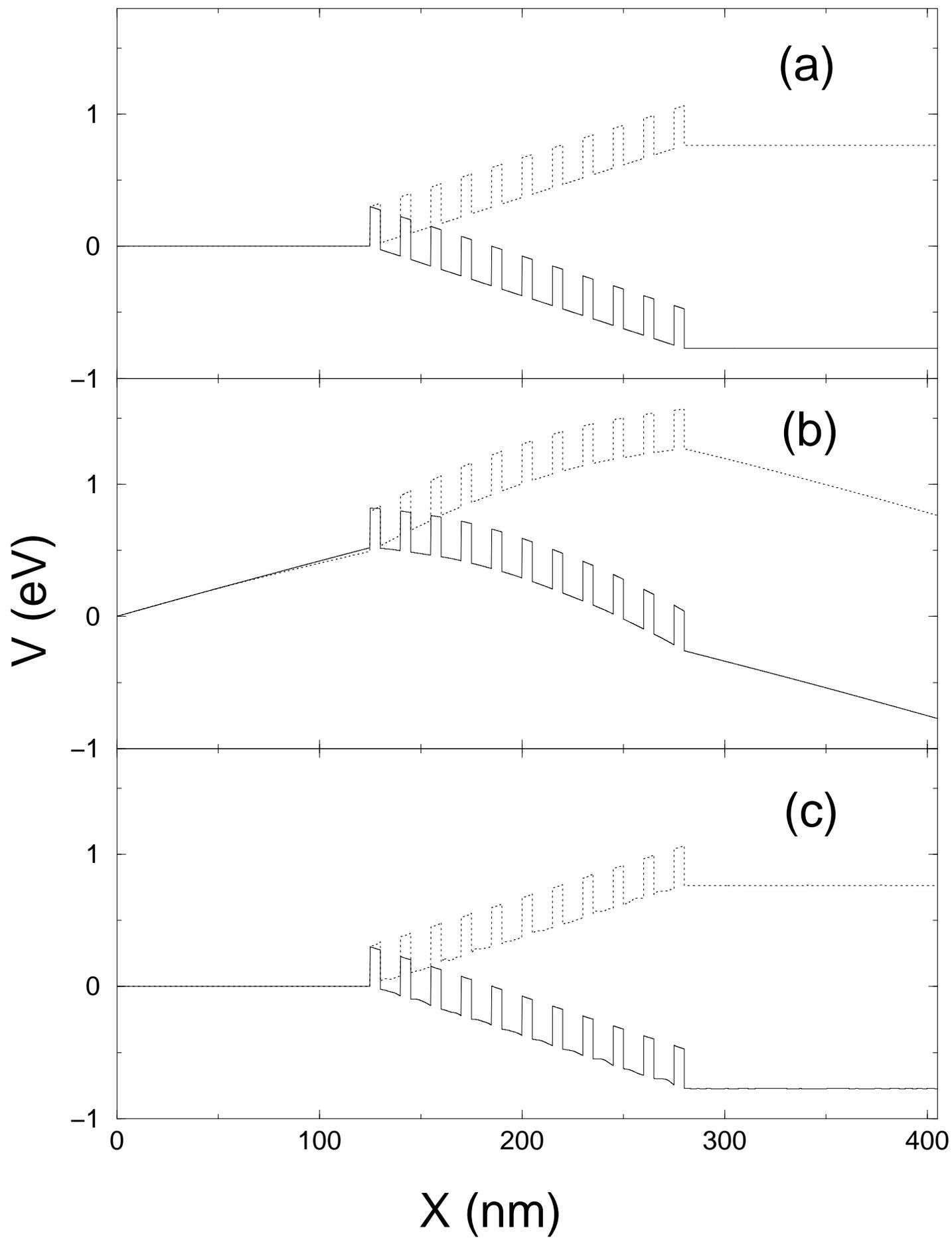
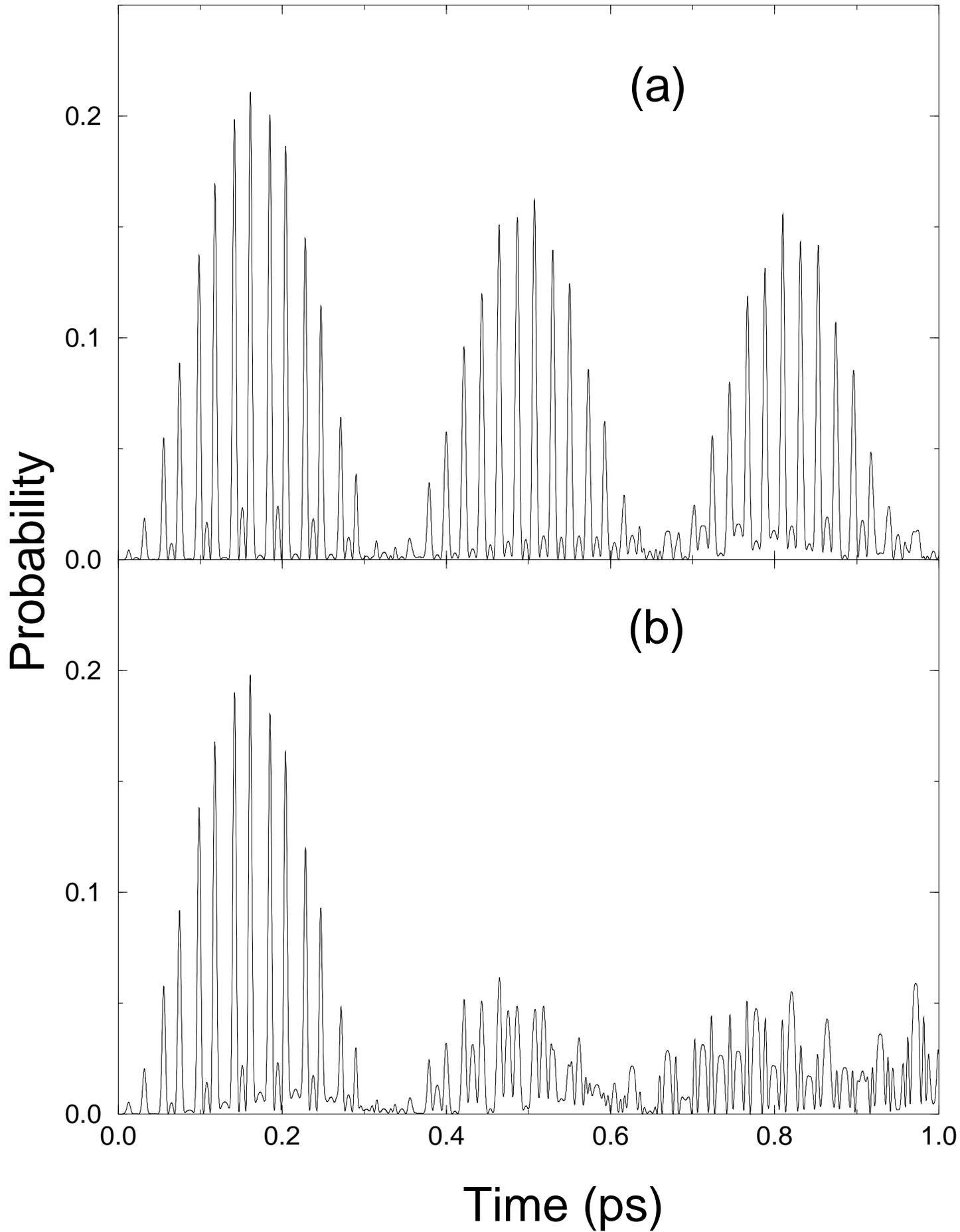
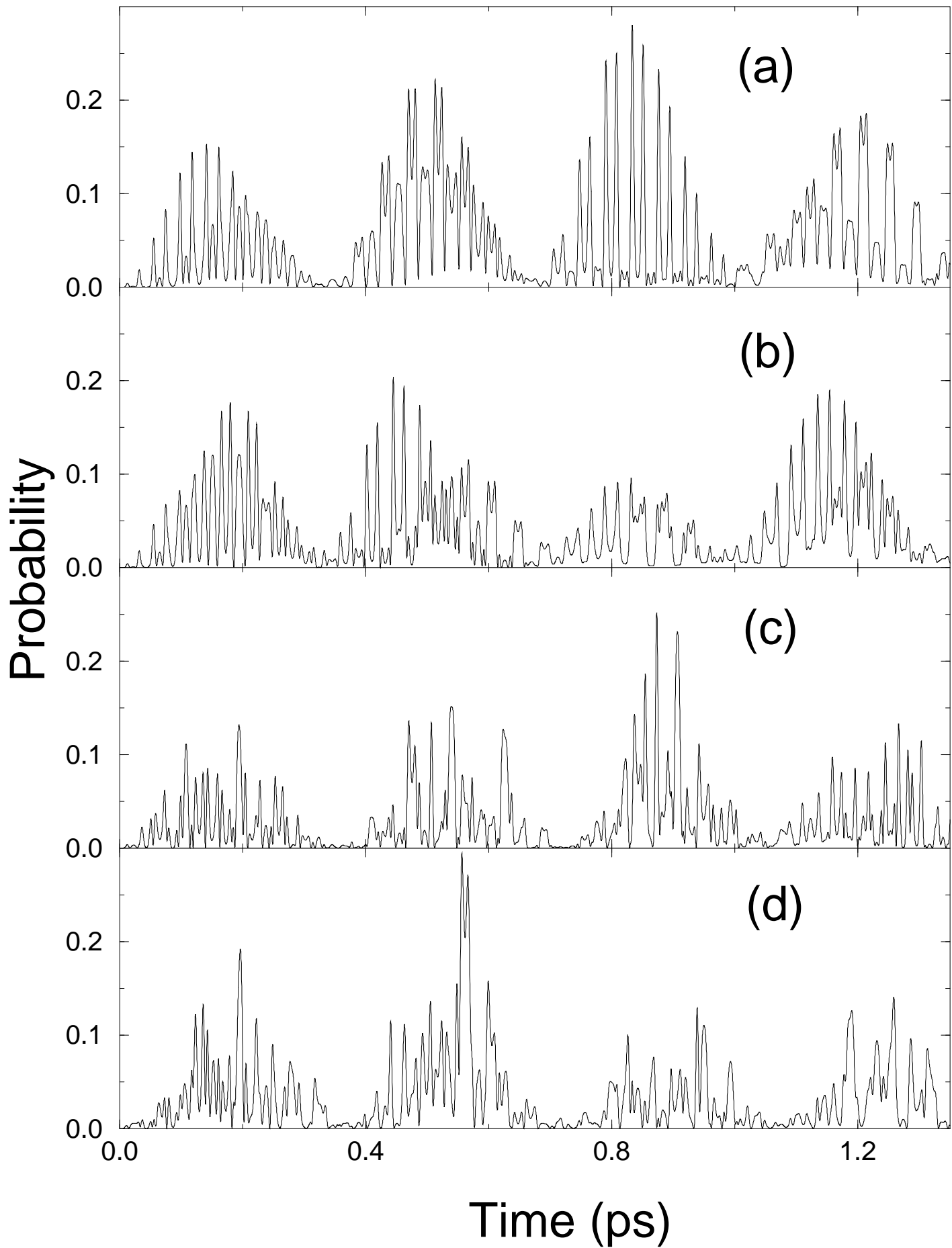
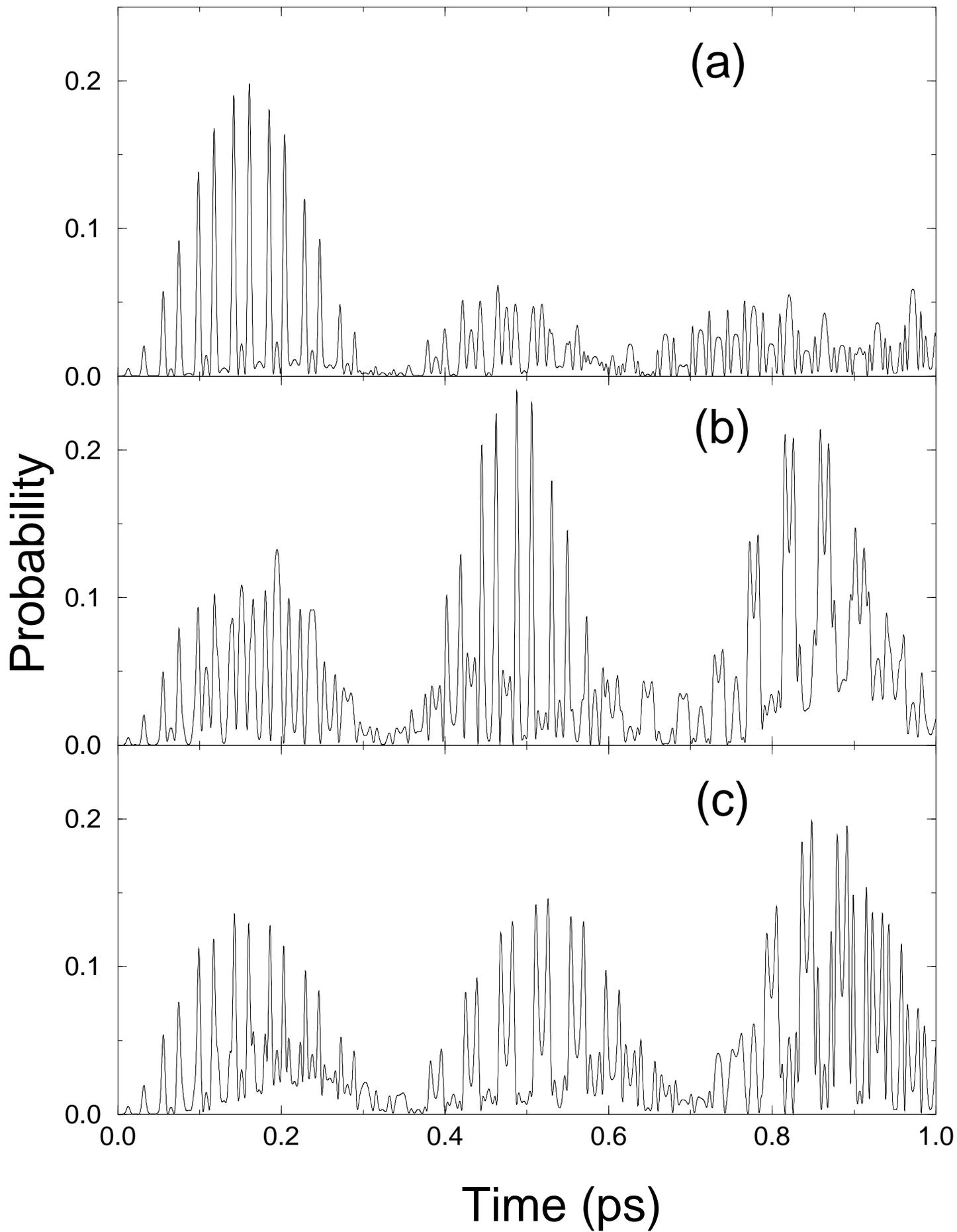
\end{document}